\documentclass[11pt,tightenlines,print,twoside,onecolumn,aps,amsmath,amssymb]{revtex4}
\usepackage{cases}
\usepackage{amsmath}
\usepackage{amssymb}
\usepackage{amsfonts}
\usepackage{amssymb}
\usepackage{dcolumn}
\usepackage{bm}
\usepackage[section]{placeins}
\usepackage{graphicx}
\usepackage{listings}
\usepackage{epstopdf}
\usepackage{tabularx}
\usepackage{hyperref}
\usepackage[caption=false]{subfig}

\usepackage{hepnicenames}
\newcolumntype{L}[1]{>{\hsize=#1\hsize\raggedright\arraybackslash}X}%
\newcolumntype{R}[1]{>{\hsize=#1\hsize\raggedleft\arraybackslash}X}%
\newcolumntype{C}[2]{>{\hsize=#1\hsize\columncolor{#2}\centering\arraybackslash}X}%

\newcommand{\be}{\begin{equation}}
\newcommand{\ee}{\end{equation}}
\newcommand{\bea}{\begin{eqnarray}}
\newcommand{\eea}{\end{eqnarray}}
\newcommand{\Fig}[1]{Fig.\,\ref{#1}}

\makeatletter
\def\tagform@#1{\maketag@@@{(\ignorespaces\textbf{#1}\unskip\@@italiccorr)}}
\renewcommand{\eqref}[1]{\textup{{\normalfont(\ref{#1}}\normalfont)}}
\makeatother

\usepackage[usenames]{color}
\definecolor{Red}{rgb}{1,0,0}
\definecolor{Blue}{rgb}{0,0,1}

\usepackage{fancyhdr}
\newcommand{\tstamp}{\today}

\begin{document}
\title{Improvement of photon reconstruction in PandoraPFA}
\author{Boruo Xu}
\affiliation{University of Cambridge}

\lhead[\fancyplain{}{\thepage}]         {\fancyplain{}}
\chead[\fancyplain{}{}]                 {\fancyplain{}{}}
\rhead[\fancyplain{}]                  {\fancyplain{}{\thepage}}
\lfoot[\fancyplain{}{}]                 {\fancyplain{\tstamp}{\tstamp}}
\cfoot[\fancyplain{\thepage}{}]         {\fancyplain{\thepage}{}}
\rfoot[\fancyplain{\tstamp} {\tstamp}]  {\fancyplain{}{}}
\begin{abstract}
This paper presents the overview of improving photon reconstruction in PandoraPFA. We have reduced the fragmentation and improved the photon separation resolution. As a result, the reconstructed photons have a greater completeness and purity, and the jet energy resolution has improved for high energy jets. 
\end{abstract}

\maketitle

Talk presented at the International Workshop on Future Linear Colliders (LCWS15), Whistler, Canada, 2-6 November 2015

\section{Introduction}{

Since the discovery of a particle consistent with being the SM Higgs boson in LHC at 2012 \cite{Aad:2012tfa,Chatrchyan:2012ufa}, our understanding of Standard Model has improved greatly. Yet limited by the underlying QCD interaction from proton-anti-proton collision, one has great difficulty to measure the properties of the Higgs precisely. Next generation electron-positron linear collider could hopefully make precision measurements of the Higgs sector and the Top quark sector \cite{Abramowicz:2013tzc}.

The leading candidates for next generation electron-positron linear collider are the International Linear Collider (ILC) \cite{Brau:2007zza}, and the Compact Linear Collider (CLIC) \cite{Linssen:2012hp}. The ILC has developed two detector models, namely the International Large Detector (ILD) \cite{Abe:2010aa} and the Silicon Detector (SiD) \cite{Aihara:2010zz}. The CLIC has developed two slightly modified detector models based on ILD and SiD \cite{Linssen:2012hp}. One key common feature of these next generation electron-positron linear colliders is the high granular calorimeter, which provides a great spatial resolution at the cost of the energy resolution. Particle flow algorithms (PFA) benefit from the spatial resolution from calorimeters, together with tracking information, to provide excellent a jet energy resolution. PandoraPFA, the most complicated and the best performing one, provides a jet energy resolution of less than 3.5\%, which is required for W/Z separation \cite{Thomson:2009rp,Marshall:2013bda}.

\begin{figure}[tbph]
\centering
{\includegraphics[width=0.5\textwidth]{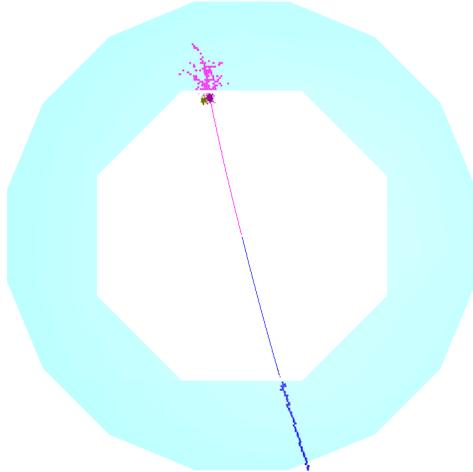}}%

\caption{An event display of a simulated $\Pem\Pep\to \Ptauon\APtauon$ event. The blue region is the cross section of the Electromagnetic Calorimeter barrel region. The top $\Ptau$ decays into a charged $\Ppi$, two photons and neutrinos. The bottom $\Ptau$ decays into a muon and neutrinos.}
\label{fig:Tautau}
\end{figure}

Photon reconstruction is an important part of particle reconstruction. For many physics processes involving particles decaying into photons, such as $\Ptau$ lepton and $\Ppizero$, a good photon reconstruction, which provides a good single photon completeness and purity, as well as a good photon separation resolution, is crucial for reconstructing these particles.

}

\section{Overview of existing photon reconstruction in PandoraPFA }{

PandoraPFA provides a framework for particle reconstruction \cite{Marshall:2015rfa}. In the linear collider content, it has a vast library of algorithms, developed through years by many people, each aiming to address one topological issue in reconstruction \cite{Thomson:2009rp,Marshall:2013bda}. The essential part is track-cluster association and reclustering to find the best track-cluster pair. Algorithms that removes clusters without tracks in the tracking detector, such as removing muon clusters or photon clusters, would provide a cleaner environment for reconstruction of charged hadrons, hence improving the jet energy resolution.

Photon identification in PandoraPFA has two main mechanisms. The basic photon identification performs photon identification test on clusters that are not matched to tracks, after track-cluster association and reclustering processes. The second photon identification, a more aggressive standalone algorithm, is performed before track-cluster association and reclustering process, aiming to remove the photon electromagnetic shower cores carefully from the rest of the complicated environment.

The addition of the second mechanism has provided an improvement for jet energy resolution, as compared to the case without it, due to correctly identifying photon electromagnetic shower cores and leaving a cleaner environment for the particle flow algorithms. However, its aggressiveness in identifying electromagnetic shower cores leaves the outer parts of the shower core as fragments, which may be reconstructed as separate particles. This undesired feature presents difficulties for anyone to use the number of reconstructed photons as a good physical quantity. Also, there is room for improvement for a better photon separation resolution, as the photon reconstruction algorithm could only identify nearby photons at four cell sizes apart in the electromagnetic calorimeter (see \Fig{fig:nPhoton}).

This paper presents a solution to the problem above, the fragmentation of photon reconstruction in PandoraPFA, as well as introducing new algorithms that improve the photon separation resolution.

The testing simulated data in this paper are generated either by WHIZARD \cite{whizard} or by the simple HepEvt generator. Events are simulated with GEANT4 \cite{Agostinelli:2002hh} in MOKKA \cite{MoradeFreitas:2002kj}. Jet fragmentation was performed with PYTHIA \cite{Sjostrand:1995iq} and the particle reconstruction was done by PandoraPFA \cite{Marshall:2015rfa} in MARLIN reconstruction framework \cite{Gaede:2006pj}, in ILD\_o1\_v6 detector model. The iLCSoft v17-01-07 was used. Different versions of PandoraPFA were used for the comparison purpose.

\section{Overview of changes to photon reconstruction}{

\begin{figure}[tbph]
\centering
\subfloat[]{{\includegraphics[width=.3\textwidth]{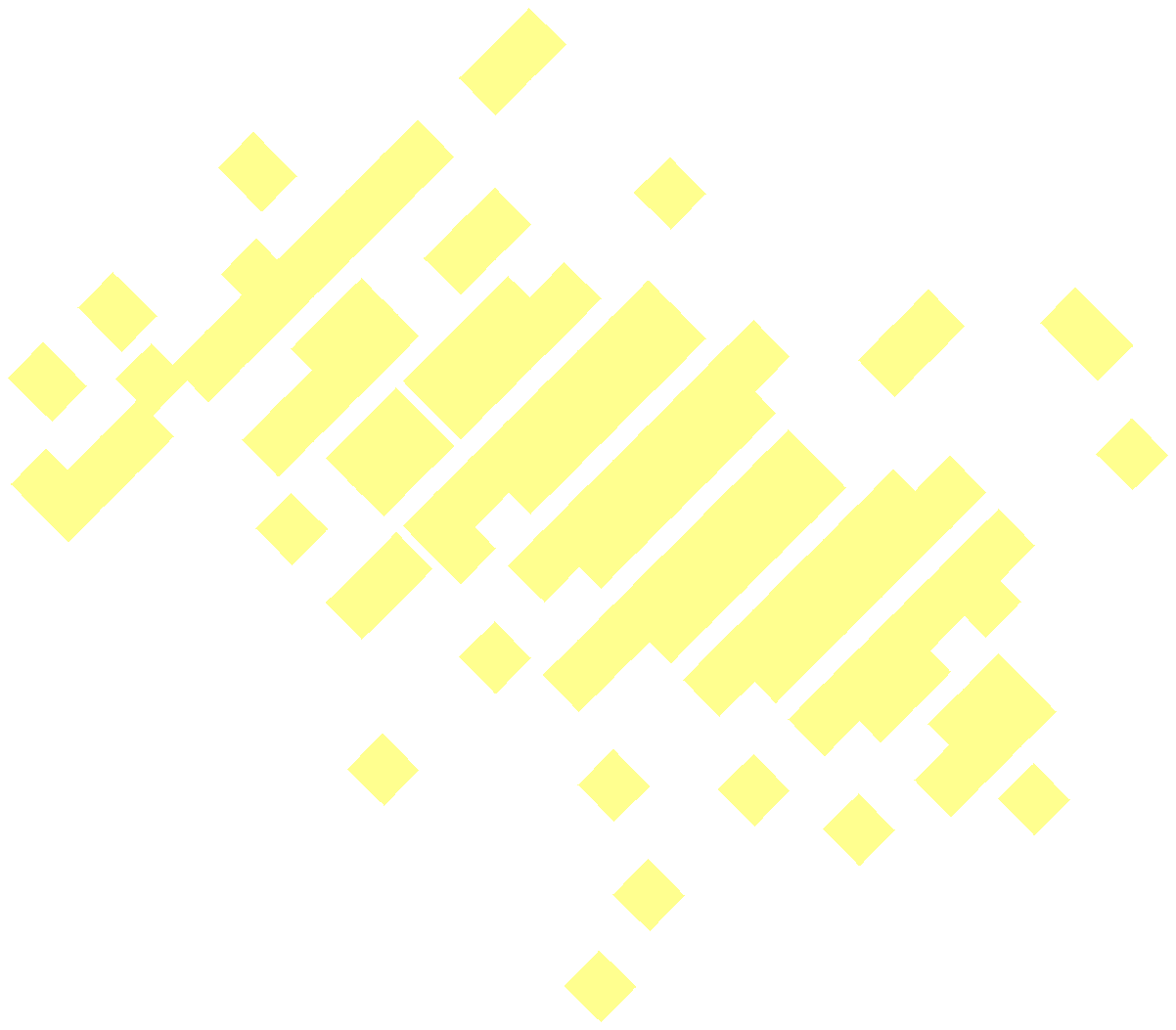} }\label{fig:evtDspPhotonFragAll}}%
\subfloat[]{{\includegraphics[width=.3\textwidth]{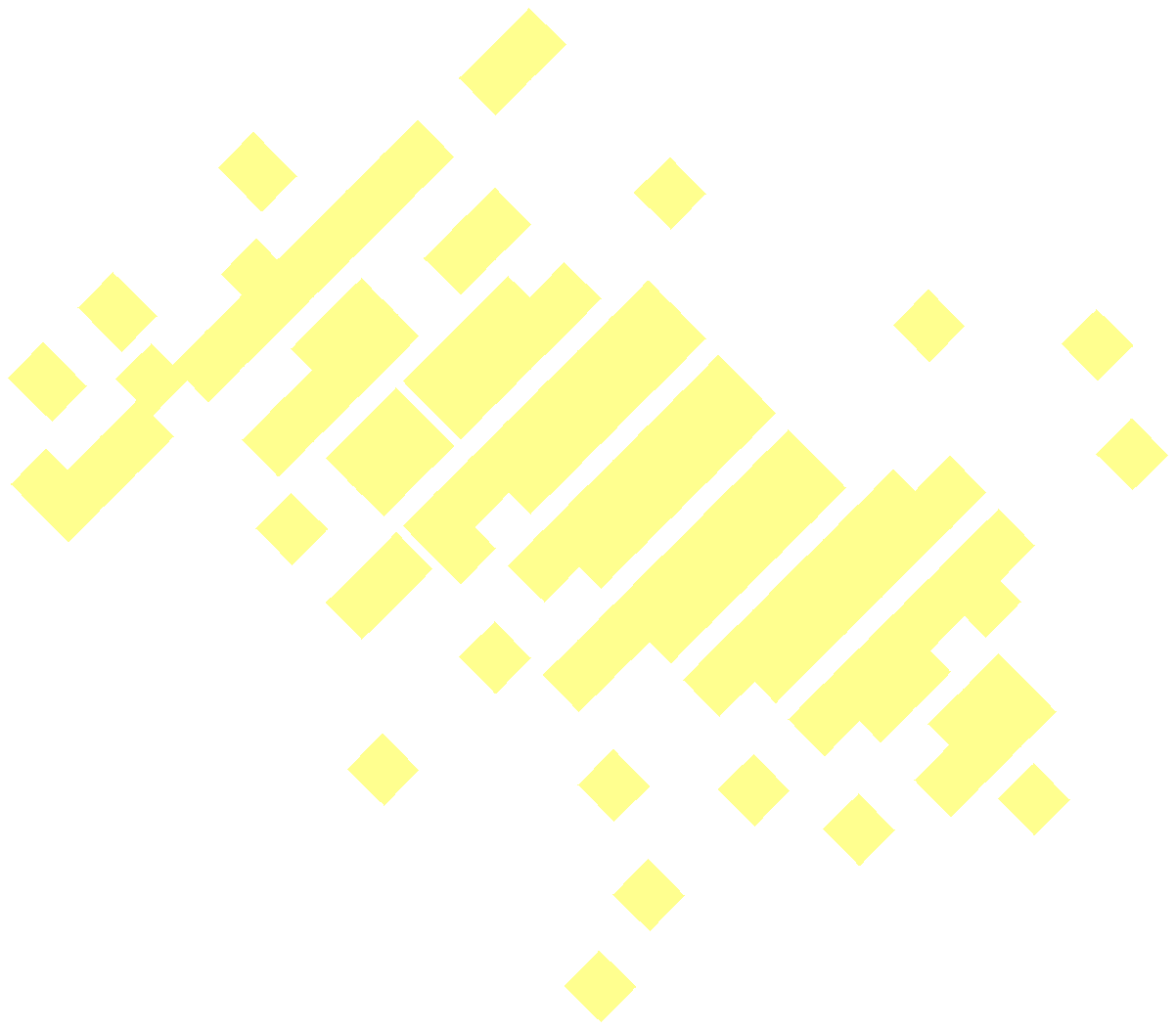} }\label{fig:evtDspPhotonFragBig}}%
\subfloat[]{{\includegraphics[width=.3\textwidth]{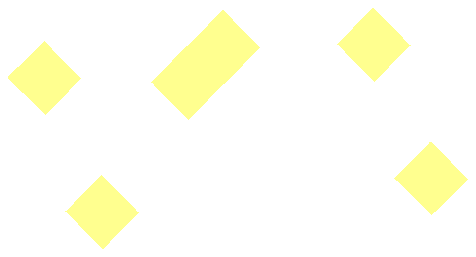} }\label{fig:evtDspPhotonFragSmall}}%
\caption{An event display of a typical 10\,GeV photon (\Fig{fig:evtDspPhotonFragAll}), reconstructed into a main photon cluster (\Fig{fig:evtDspPhotonFragBig}) and a photon fragment cluster (\Fig{fig:evtDspPhotonFragSmall}). }
\label{fig:evtDspPhotonFrag}
\end{figure}

\Fig{fig:evtDspPhotonFrag} illustrates a typical creation of photon fragments in the electromagnetic calorimeter (ECal). When the core of the photon electromagnetic shower is identified as a photon (the main photon), the outer part of the shower is reconstructed as a separate particle, and wrongly identified as a photon or a neural hadron (the photon fragment). The photon fragment does not have the typical shower like structure of a main photon, and typically it is has much lower energy than the main photon. Thus these characteristics of fragments provide a way of merging photon fragments to the main photon.

To merge photon fragments to the main photon, we carefully compare distributions of various quantities for main photon-fragment pairs and main photon-other particles pairs, and choose cuts that would merge fragments to main photons and not merge other particles to main photons. The most useful distances related parameters are the simple distance between centroids of the pair, and the mean intra-layer distance weighted over energy between the pair, where the main photon-fragment pairs should have a small distance metric and main photon-non-photon particles pairs would have larger ones. The other useful method is the energy deposition of the pair in a transverse plane perpendicular to the direction of the flight, where the main photon-fragment pairs produce a single peak in the two dimensional shower, and the main photon-non-photon particles pairs produce two peaks in the two dimensional shower.

Based on this, two algorithms have been carefully implemented to merge fragments in electromagnetic calorimeter to the main photons. One algorithm (RecoPhotonMergingAlgorithm) is placed immediately after the standalone photon identification. The other one (PhotonMergingAlgorithm) is placed at the end of the reconstruction chains in PandoraPFA. The two algorithms provide an excellent reduction in the number of fragments created, as shown in \Fig{fig:nPhoton}.

\begin{figure}[tbph]
\centering
{\includegraphics[width=0.5\textwidth]{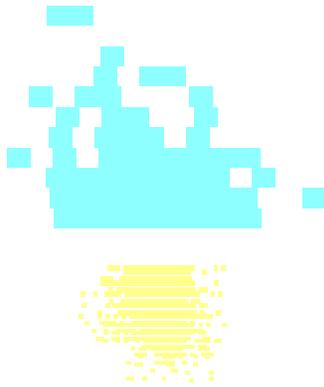}}%

\caption{An event display of a typical 500\,GeV photon, reconstructed into a main photon cluster in yellow and a neutral hadron fragment in blue, formed by energy deposition in hadronic calorimeter.}
\label{fig:evtDspHCalFrag}
\end{figure}

Another type of photon fragment is caused when a highly energetic photon, typically more than 50GeV, goes through the ECal and deposits energy in the hadronic calorimeter (HCal), illustrated in \Fig{fig:evtDspHCalFrag}. Since the photon identification in PandoraPFA implicitly assumes that all photons are contained in the ECal, the energy deposited in the HCal either becomes neutral hadron fragments or part of a hadronic shower if a hadronic shower of another particle is nearby.

Similar to the treatment of fragments in the ECal, we merge the fragments in the HCal by calculating various quantities and place selection cuts. The important quantities are the distance between centroid for main photon and fragments, the distance of closest approach of the directions of the flights for main photon and fragments, and the fraction of energy in a fitted cone in HCal by extending the fitted cone from main photon in the ECal, where for main photon-fragment pairs, the distance metrics are small and the fraction of the energy in a fitted cone in HCal are large.

We implemented the algorithm in PandoraPFA (HighEnergyPhotonRecovery), after the standalone photon reconstruction algorithm, which successfully remove most of photon fragments in the HCal for high energy photons, as shown in \Fig{fig:n_all}.

\begin{figure}[tbph]
\centering
{\includegraphics[width=0.5\textwidth]{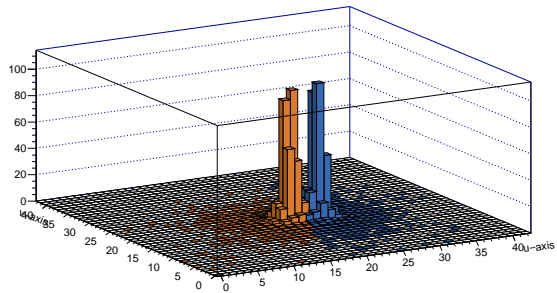}}%

\caption{Two 500\,GeV photons (yellow and blue), being just resolved in the transverse plane perpendicular to the direction of the flight, of their energy deposition in electromagnetic calorimeter. U and V axis are two arbitrary axis perpendicular to each other in the plane. Z axis is the sum of the calorimeter hit energy in each particular bin in 2D plane in GeV.}
\label{fig:peakFinding}
\end{figure}

With these new fragment merging algorithms, we have removed most of the photon fragments. We also improved the photon separation resolution by rewriting the existing standalone photon reconstruction algorithm and the existing 2D peak finding algorithm.

The core part of the standalone photon reconstruction algorithm is to identify peaks in a 2D plane, the energy deposition of a transverse 2D plane perpendicular to the direction of the flight. Each identified peak will form a cluster. The cluster will be identified as a photon if it passes the photon likelihood test. The key to improve photon separation resolution is to improve the 2D peak finding algorithm. We have implemented a new peak finding algorithm, which finds all possible local maxima. Other cells in the 2D plane are associated to each peak by minimising the metric $d/\sqrt(E)$, where $d$ is the distance in 2D plane from the cell to the peak, and $E$ is the energy or height of that particular peak. \Fig{fig:peakFinding} shows two energetic photons just resolved with the new 2D peak finding algorithm.

We also carefully identify photons close to charged hadrons, by discarding identified peaks in 2D plane that are associated with tracks. The likelihood test for photons is improved by taking the normalisation factor into account.

As shown in \Fig{fig:JER}, jet energy resolution at high energy has improved due to the improving ability to identify photons close to charged hadrons.

We also implemented a new photon separation algorithm (PhotonSplittingAlgorithm) at the end the reconstruction in PandoraPFA to separate photons that are wrongly merged into one photon by algorithms in the construction chain. The photon separation algorithm uses the new 2D peak finding algorithm and separate photon clusters according to peaks in the 2D plane. Hence if two photons are merged and are identified as two peaks in the 2D plane, the new algorithm would correctly separate the merged photon into two photons. As shown in \Fig{fig:n_p}, the separation of two photons improved from previously four cell size as minimum separation distance to two cell size.

}

\section{Figures of merit}{
\begin{figure}[tbph]
\centering
\subfloat[]{{\includegraphics[width=.5\textwidth]{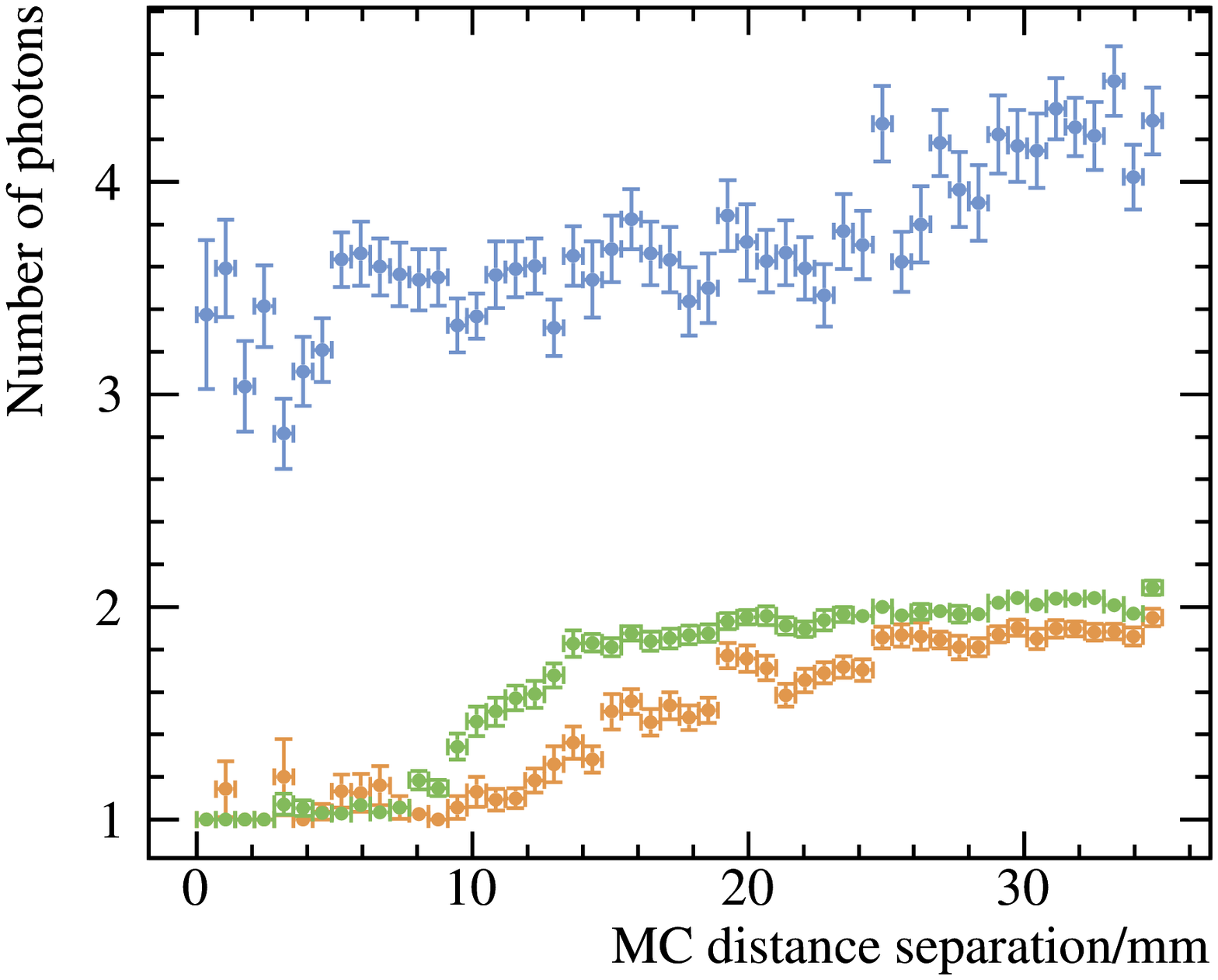} }\label{fig:n_p}}%
\subfloat[]{{\includegraphics[width=.5\textwidth]{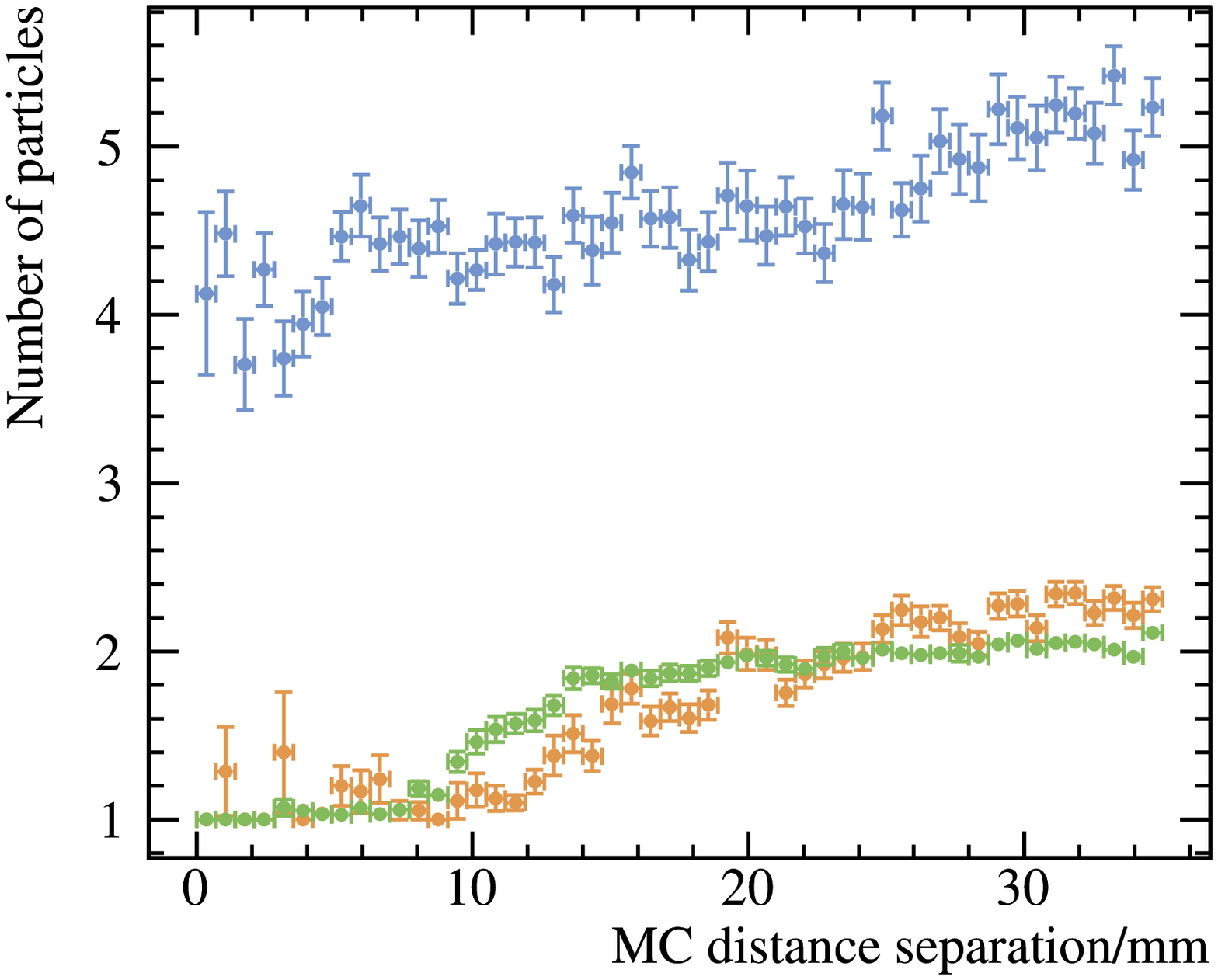} }\label{fig:n_all}}%

\caption{\Fig{fig:n_p} shows the number of reconstructed photons as a function of their true distance separation for a two photons per event sample. \Fig{fig:n_p} shows the number of reconstructed particle as a function of their true distance separation for a two photons per event sample. The top blue line is reconstructed with PandoraSettingsDefault.xml with iLCSoft v01-17-07. The bottom orange line is reconstructed with PandoraSettingsDefault.xml plus two fragment merging algorithm (RecoPhotonFragmentMergingAlgorithm and PhotonFragmentMergingAlgorithm). The middle green line is is reconstructed with PandoraSettingsDefault.xml plus all algorithms described above.
}
\label{fig:nPhoton}
\end{figure}

We will review performance metrics of above algorithms. \Fig{fig:n_p} shows the number of reconstructed photons as a function of their true distance separation for a two photons per event sample. The reduction of the number of reconstructed photons are mainly due the the fragment merging algorithms for fragments in the ECal. \Fig{fig:n_all} shows a similar reduction in the reconstructed particles as in \Fig{fig:n_p}, and it shows that neutral hadron fragments in HCal have been merged back to main photons.

\Fig{fig:nPhoton} also shows the separation resolution is improved with new 2D peak finding algorithms and new photon reconstruction. Two photons start to be separated at a much shorter distance, typically starting to resolve when they are two cell size (10\,mm) apart comparing to previously four cell size (20\,mm) apart.

\begin{figure}[tbph]
\centering
{\includegraphics[width=.5\textwidth]{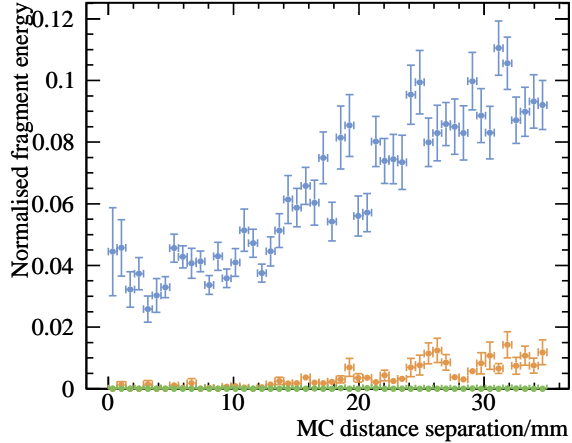}}

\caption{Figure shows the normalised energy in the fragment as as a function of their true distance separation for a two photons per event sample. The normalised fragment energy is defined as the energy in fragments divided by the total reconstructed energy in the event. The blue line is reconstructed with PandoraSettingsDefault.xml with iLCSoft v01-17-07. The orange line is reconstructed with PandoraSettingsDefault.xml plus two fragment merging algorithm (RecoPhotonFragmentMergingAlgorithm and PhotonFragmentMergingAlgorithm). The green line is is reconstructed with PandoraSettingsDefault.xml plus all algorithms described above.
}
\label{fig:fragEnergy}
\end{figure}

\Fig{fig:fragEnergy} shows the energy in the fragment as as a function of their true distance separation for a two photons per event sample. The fraction of energy in the photon fragments has reduced considerably, which is consistent with the reduction of number of reconstructed photons. The green line which is shows the best performance, comparing with the orange line which did not contains the PhotonSplittingAlgorithm and the new 2D Peak Finding algorithm, indicates the fragments in the HCal are merged to the main photon correctly.

\begin{figure}[tbph]
\centering
{\includegraphics[width=.5\textwidth]{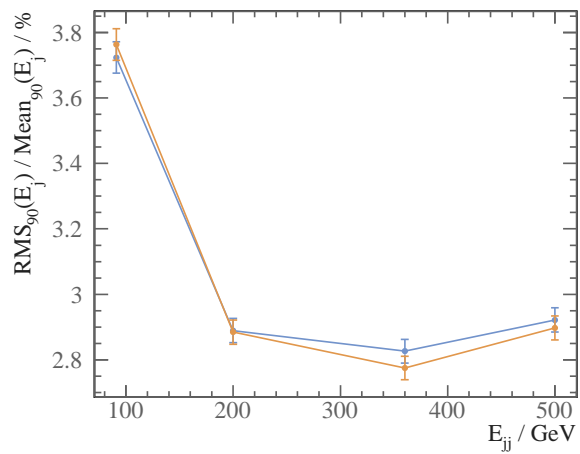}}

\caption{Figure shows jet energy resolution as a function of the di-jet energy. The sample is Z\' decay u/d/s. The jet energy resolution is defined as the root mean squared divided by the mean for the smallest width of distribution that contains 90\% of entries. The top blue line is reconstructed with PandoraSettingsDefault.xml with iLCSoft v01-17-07. The bottom orange line is is reconstructed with PandoraSettingsDefault.xml plus all algorithms described above.
}
\label{fig:JER}
\end{figure}

\Fig{fig:JER} shows the comparison of jet energy resolution for di-jet energy of 91, 200, 360 and 500\,GeV, before and after the improvement. Although there is a small degradation at low energy, the jet energy resolution improves at high energy.

}

\section{Conclusion }{

This paper presents the algorithms for improving photon reconstruction in PandoraPFA, reducing the fragmentation and to improve the photon separation resolution. As a result, reconstructed photons have a greater completeness and purity, and the jet energy resolution has improved for high energy jets.

This will feed back to physics studies that involves photons as part of the final decay states. For example, $\Ptau$ lepton and $\Ppizero$ decay and single photon study would benefit from the improved photon reconstruction.

Future work could consist of improving the photon separation resolution further by using template fit for two photons that are inseparable by 2D peak finding algorithms.

The authors appreciate J. S. Marshall and M. A Thomson for discussions and helpful comments.

}

\bibliographystyle{h-physrev3}
\bibliography{bibliography}

\enddocument}
\end